\documentstyle[preprint,aps]{revtex}

\begin{document}
\title{The phase diagram of the one-dimensional quantum sine-Gordon system ($%
\beta ^{2}=4\pi )$ with a linear spatial modulation }
\author{Zhiguo Wang, Yumei Zhang}
\address{Department of Physics, Tongji University, Shanghai, 200092, China}
\date{\today}
\maketitle

\begin{abstract}
The one-dimensional quantum sine-Gordon system with a linear spatial
modulation is investigated in a special case, $\beta ^{2}$ =4$\pi $. The
model is tranformed into a massive Thirring model and then is exactly
diagonalized, the energy spetrum of the model is obtained. Our result
clearly demonstrates that cancelling the cosine term without any considering
is unadvisable.
\end{abstract}

The one-dimensional (1D) quantum sine-Gordon model$^{\text{\cite{Sklyanin}}}$
may probably be the most useful quantum model since it can be used to
describe the most of the one- and two-dimensional (2D) models of either
fermi or bose system,$^{\text{\cite{Minnhagen}},\text{\cite{E.Fradkin},\cite%
{Gogolin}}}$ this fact attaches particular importance to the quantum
sine-Gordon model. Many works have been done about this model both in field
theory and in condensed matter physics.$^{\text{\cite{Sklyanin}},\text{\cite%
{Coleman}},\text{\cite{Ingermanson}},\text{\cite{Mandelstam}},\text{\cite%
{Stevenson}}}$ It is exactly solvable by quantum inverse scattering method.$%
^{\text{\cite{Sklyanin}}}$ By a variational method Coleman$^{\text{\cite%
{Coleman}}}$ first discovered that the energy density of the system is
unbounded below when the coupling constant $\beta ^{2}$ exceeds $8\pi $, so
there is a phase transition as the coupling constant varies. This
corresponds to the Kosterlitz-Thouless (K-T) phase transition by the
equivalence of the 2D Coulomb gas and sine-Gordon model. The soliton mode of
the sine-Gordon model corresponds to a one-fermion excitation in the fermi
picture, which was clarified soon later by Mandelstam by introducing a
Fermi-Bose relation.$^{\text{\cite{Mandelstam}}}$

As there are other spatial variations in the cosine potential, the low
energy properties of the 1D quantum sine-Gordon system are more difficult to
be analysized. Here we shall discuss a simple case that there are a linear
spatial modulation in the cosine potential. The Hamlitonian reads as 
\begin{equation}
H=\int \left\{ \frac{1}{2}\left[ \Pi ^{2}(x)+\left( \frac{\partial \phi (x)}{%
\partial x}\right) ^{2}\right] -\frac{\alpha }{\beta ^{2}}\cos \left( \beta
\phi (x)+\lambda x\right) \right\} dx\text{ ,}
\end{equation}%
here $\phi (x)$ is a bose field operator, $\Pi (x)=-i\frac{\delta }{\delta
\phi (x)}$ is its conjugate momentum, they satisfy the commutation relation 
\begin{equation}
\left[ \phi (x),\Pi (y)\right] =i\delta (x-y)\text{ . }
\end{equation}%
$\lambda $ is a spatial modulated parameter. In condensed matter physics $%
\lambda $ represents the fermi surface shifting from half filling in the
fermi picture.$^{\text{\cite{Sun}}}$

In the presence of finite $\lambda $, most previous works suggested that the
above Hamlitonian (1) describes a massless free field and cancelled the
cosine potential directly. Has the cosine potential really no effect on the
system in this case?

Schulz discussed a one-dimensional quantum sine-Gordon system with an
additional gradient term.$^{\text{\cite{Schulz}}}$ If we take a shift, $\phi
+\frac{4m\pi x}{\beta a}\rightarrow \phi $, the above Hamiltonian (1) is
same as that of Schulz. But if one pays attention to the boundary condition
of bose operator $\phi $, he will find both model are different since the
boundary conditions of system will be altered under this shift, namely, $%
\int_{0}^{L}\nabla \phi (x)dx=0\rightarrow \int_{0}^{L}\nabla \phi (x)dx=%
\frac{\lambda }{\beta }L$, therefore the eigenstates will be also changed.
Schultz pointed out that a commensurate-incommensurate transition happens in
case of finite coefficient of the gradient term, this result implies that
directly cancelling the cosine potential is unsuited.

In order to find an unquestionable answer for this, we investigate this
Hamlitonian in a special case, $\beta ^{2}$ =4$\pi $. First we transform the
model into the massive Thirring model by the operator identities between
fermions and bosons, and then exactly diagonalize the model using the
bogliubov transformation. When the cosine potential may be omitted is
obvious in our results.

For the case of a finite $\lambda $, the above system has rarely been
discussed before. After we use the bose-fermi relations$^{\text{\cite{Takada}%
}}$%
\[
\frac{1}{2}\int :\Pi ^{2}(x)+\left( \frac{\partial \phi (x)}{\partial x}%
\right) ^{2}:dx=\text{ \ \ \ \ \ \ \ \ \ \ \ \ \ \ \ } 
\]%
\begin{equation}
-i\int \left[ \psi _{1}^{\dagger }(x)\frac{\partial }{\partial x}\psi
_{1}(x)-\psi _{2}^{\dagger }(x)\frac{\partial }{\partial x}\psi _{2}(x)%
\right] dx\text{ ,}
\end{equation}%
\begin{equation}
\cos \left( 2\sqrt{\pi }\phi (x)\right) =\pi \epsilon \left[ \psi
_{1}^{\dagger }(x)\psi _{2}(x)+\psi _{2}^{\dagger }(x)\psi _{1}(x)\right] 
\text{ ,}
\end{equation}%
the Bose Hamlitonian (1) in the case $\beta ^{2}$ =4$\pi $ can be
transformed into a modified massive Thirring model%
\begin{eqnarray}
H &=&\int \left\{ -i\left[ \psi _{1}^{\dagger }(x)\frac{\partial }{\partial x%
}\psi _{1}(x)-\psi _{2}^{\dagger }(x)\frac{\partial }{\partial x}\psi _{2}(x)%
\right] \right. \text{ }  \nonumber \\
&&-\left. \frac{\alpha \epsilon }{4}\left[ e^{i\lambda x}\psi _{1}^{\dagger
}(x)\psi _{2}(x)+e^{-i\lambda x}\psi _{2}^{\dagger }(x)\psi _{1}(x)\right]
\right\} dx\text{ .}
\end{eqnarray}%
Here $\epsilon $ is an infinitesimal positive parameter, it is of the order
of the lattice constant in condensed matter physics. With following Fourier
transformations 
\begin{eqnarray}
c_{1k} &=&\frac{1}{\sqrt{L}}\int \psi _{1}(x)e^{i(k-\frac{\lambda }{2})x}dx%
\text{ ,}  \nonumber \\
\text{ }c_{2k} &=&\frac{1}{\sqrt{L}}\int \psi _{2}(x)e^{i(k+\frac{\lambda }{2%
})x}dx\text{ ,}
\end{eqnarray}%
the Hamlitonian (5) is rewritten as%
\begin{equation}
H=\sum_{k}\left[ (k-\frac{\lambda }{2})c_{1k}^{\dagger }c_{1k}-(k+\frac{%
\lambda }{2})c_{2k}^{\dagger }c_{2k}\right] -\frac{\alpha \epsilon }{4}%
\sum_{k}\left[ c_{1k}^{\dagger }c_{2k}+c_{2k}^{\dagger }c_{1k}\right] \text{
,}
\end{equation}%
where $\sum $ denotes the summation over momentum which is cut off at $%
\Lambda (\sim 1/\epsilon )$.

In order to diagonalize one particle terms we apply a Bogolubov
transformation to the Hamlitonian (7)%
\begin{eqnarray}
c_{1k} &=&u_{k}\beta _{k}+v_{k}\alpha _{k}^{\dagger }\text{ ,}  \nonumber \\
c_{2k} &=&u_{k}\alpha _{k}^{\dagger }-v_{k}\beta _{k}\text{ .}
\end{eqnarray}%
The standard technique gives us a quasi-particle Hamlitonian%
\begin{equation}
H=\sum_{k}\left[ (\sqrt{k^{2}+\left( \frac{\alpha \epsilon }{4}\right) ^{2}}-%
\frac{\lambda }{2})\beta _{k}^{\dagger }\beta _{k}+(\sqrt{k^{2}+\left( \frac{%
\alpha \epsilon }{4}\right) ^{2}}+\frac{\lambda }{2})\alpha _{k}^{\dagger
}\alpha _{k}\right] \text{ .}
\end{equation}

The quasi-particle creation operator $\beta _{k}^{\dagger }$ and $\alpha
_{k}^{\dagger }$ correspond to two different quasi-particles such as the
hole or particle. But now the excitations of them is not same. When $\alpha
\epsilon >2\lambda $, both have excitation gap, while the excitation of $%
\beta _{k}^{\dagger }$ is gapless in the case $\alpha \epsilon <2\lambda $.
This shows that the original model can be treated as massless free system
with spatial modulation when the parameters satisfy%
\begin{equation}
\alpha \epsilon <2\lambda \text{ .}
\end{equation}%
Otherwise, the original model can not be treated as a massless system. In \
this case, one should be always aware of that the micro-structure of the
ground state is essentially different with that of $\lambda =0$ case.

In summary, we have concluded that the cosine potential term can be
cancelled ,i.e., the model can be considered as a massless free system, when
the parameters satisfy equation (10). Although we only discussed in a
special case that $\beta ^{2}=4\pi $, it is sufficient for showing that
cancelling the cosine potential term is unadvisable without considering the
varying of parameters.

\end{document}